%% file: b2kstomega_text.tex
\documentclass[prl,aps,twocolumn,tightenlines,superscriptaddress]{revtex4}

\input epsf
\usepackage{tabularx}
\usepackage{epsfig}
\usepackage{verbatim}
\usepackage{subfigure}
\usepackage{graphicx}
\usepackage{amssymb}
\usepackage{epstopdf}
\usepackage{color}
\usepackage{xspace}
\usepackage[amssymb,pstricks,Gray,squaren]{SIunits}
\usepackage{feynmf}
\usepackage{header}
\usepackage{graphics}
\usepackage{color}

\RequirePackage{relsize} 
\begin{document}

\title{{\boldmath 
\begin{flushright}
{\small BELLE-CONF-0750}\\
{\small UCHEP-07-05}
\end{flushright}
\vskip0.10in
Search for \B Meson Decays to \OmeKstz }}

\input{author-conf2007.tex}
\noaffiliation

\begin{abstract}
We report a search for the charmless vector-vector decay \BzToOmeKstz with \NBBbar \BBbar pairs collected with the Belle detector at the KEKB \ee collider. We measure the branching fraction in units of ${10^{-6}}$: \brfrnoexp, where the first error is statistical, the second systematic, and the upper limit is at the 90\% confidence level.
\end{abstract}

\maketitle

\begin{figure}[b]
\centering
\vspace{7mm}
\subfigure{ \input{diagram.tex}} \\
\vspace{1mm}
\caption[]{\label{feynman_diagram} Penguin diagram for \BzToOmeKstz decays.}
\end{figure}
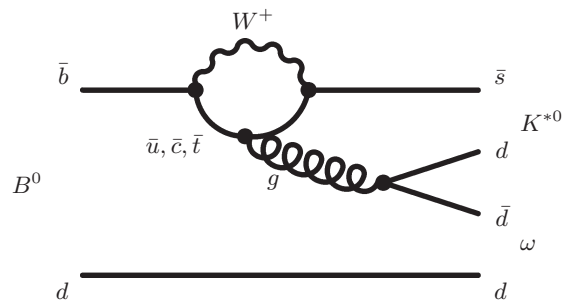

Recently, \bsqq penguin decays have received much attention in the literature. These decays proceed via an internal loop diagram and thus are potentially sensitive to new types of propagators and couplings. Such decays have sometimes yielded unexpected results, e.g., the \bsuu decay \BzToKpPim exhibits substantial direct $CP$ violation~\cite{belle_kpiCPV,babar_kpiCPV}, and the \bsss decay \BToPhiKst exhibits large transverse polarization~\cite{belle_phiKstz,babar_phiKstz}. This latter observation implies that non-factorizable contributions to the decay amplitude play a significant role. Here we search for the \bsdd decay \BzToOmeKstz (Fig. \ref{feynman_diagram}), which has not yet been observed~\cite{babar_wKstz_1,babar_wKstz_2}. The expected standard model (SM) rate is small~\cite{br_theory}, and observing an enhancement above this rate could indicate new physics. Furthermore, \BzToOmeKstz decays can be useful for determining the Cabibbo-Kobayashi-Maskawa (CKM)~\cite{ckm} angle ${\phi_{3} (= \gam)}$~\cite{phi3}.

This analysis uses \dataset of data containing \NBBbar \BBbar pairs. The data was collected with the Belle detector~\cite{belle_detector} at the KEKB~\cite{kekb} \ee asymmetric-energy (3.5 GeV on 8.0 GeV) collider with a center-of-mass (CM) energy at the \ups resonance. The production rates of \BzBzbar and \BpBm pairs are assumed to be equal. 

The Belle detector is a large-solid-angle spectrometer. It consists of a silicon vertex detector (SVD), a 50-layer central drift chamber (CDC), an array of aerogel threshold  Cherenkov counters (ACC), time-of-flight scintillation counters (TOF), and an electromagnetic calorimeter comprised of CsI(Tl) crystals located inside a superconducting solenoid coil that provides a 1.5 T magnetic field. An iron flux return located outside the coil is instrumented to detect $K^0_L$ mesons and to identify muons (KLM).

The \B-daughter candidates are reconstructed through their decays \decayOme, \decayKstz and \decaypiz~\cite{charge_conjugate_text}. A charged track is identified as a pion or kaon by combining information from the CDC, ACC and TOF systems. We reduce the number of poor quality tracks by requiring that ${|dz| < 4.0~\cm}$ and ${dr < 0.2~\cm}$, where ${|dz|}$ and ${dr}$ are the closest approach of a track to the interaction point in the $z$-direction and in the transverse plane, respectively. In addition, we require that each charged track have a transverse momentum ${p_{T} > 0.1~\gevc}$ and a minimum number of SVD hits. Tracks matched with clusters in the ECL that are consistent with an electron hypothesis are rejected.

Candidate \piz mesons are reconstructed from pairs of photons, where the energy of each photon in the laboratory frame is required to be greater than 100 MeV for the ECL endcap regions (\ECLendcapLow or \ECLendcapHigh) and 50 MeV for the ECL barrel region (\ECLbarrel), where ${\theta}$ denotes the polar angle with respect to the beam axis. We select \piz mesons with an invariant mass in the range \pizcut and a momentum in the laboratory frame \pizmomcut.  

We define \Ome and \Kstz signal regions \Omefit and \Kstzsig, respectively. In the maximum-likelihood (ML) fit described below, the \Kstz fit region extends to \Kstzfit to allow for greater discrimination between signal \BzToOmeKstz and non-resonant \BzToOmeKpPim decays. To reduce combinatorial background arising from low-momentum kaons and pions, we require that \Kstzhelcut, where \Kstzhel is the angle between the direction of the \kp and the direction opposite to the \Bz momentum in the \Kstz rest frame. 

Candidate \BzToOmeKstz decays are identified using the energy difference (\de) and the beam-energy-constrained mass (\mbc). They are defined as \defde and \defmbc, where \ebeam denotes the beam energy and \eb and \pb denote the energy and momentum, respectively, of the candidate \B-meson,  all evaluated in the \ee CM frame. We select events satisfying \defit and \mbcfit, and define signal regions \desig and \mbcsig.

\begin{figure}[t]
\mbox{\epsfxsize=1.68in \epsfbox{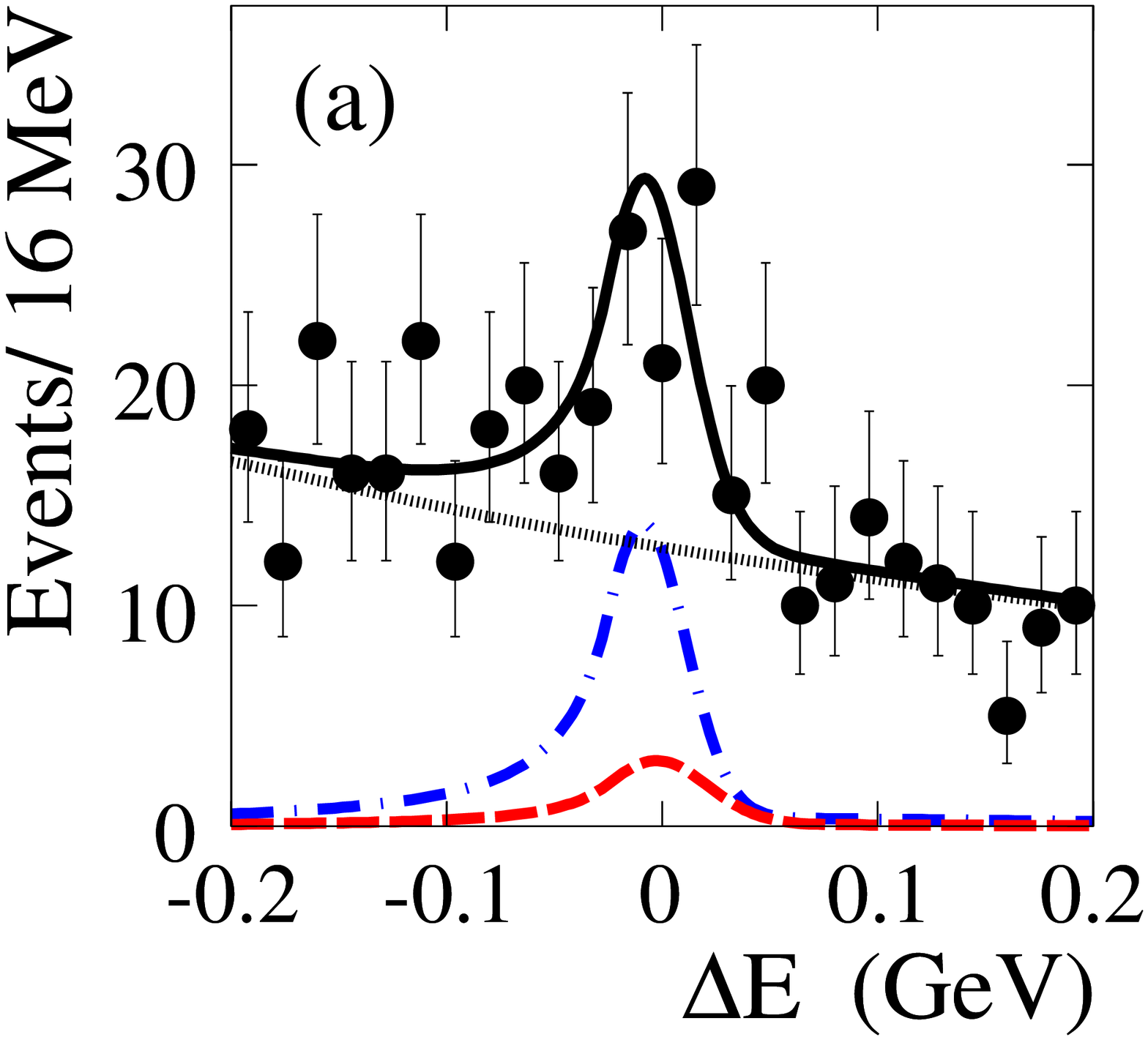}}
\mbox{\epsfxsize=1.68in \epsfbox{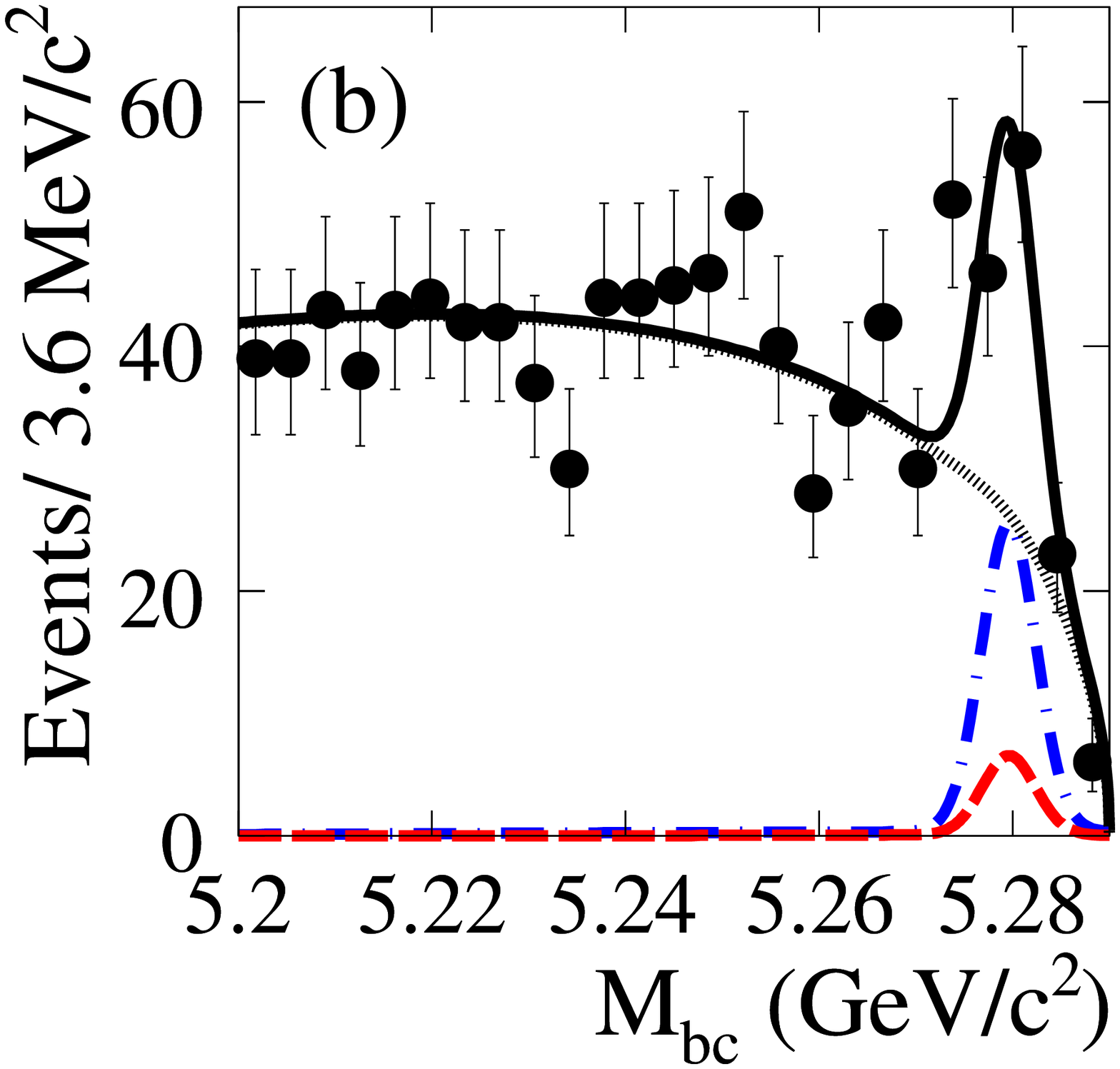}}\\
\mbox{\epsfxsize=1.68in \epsfbox{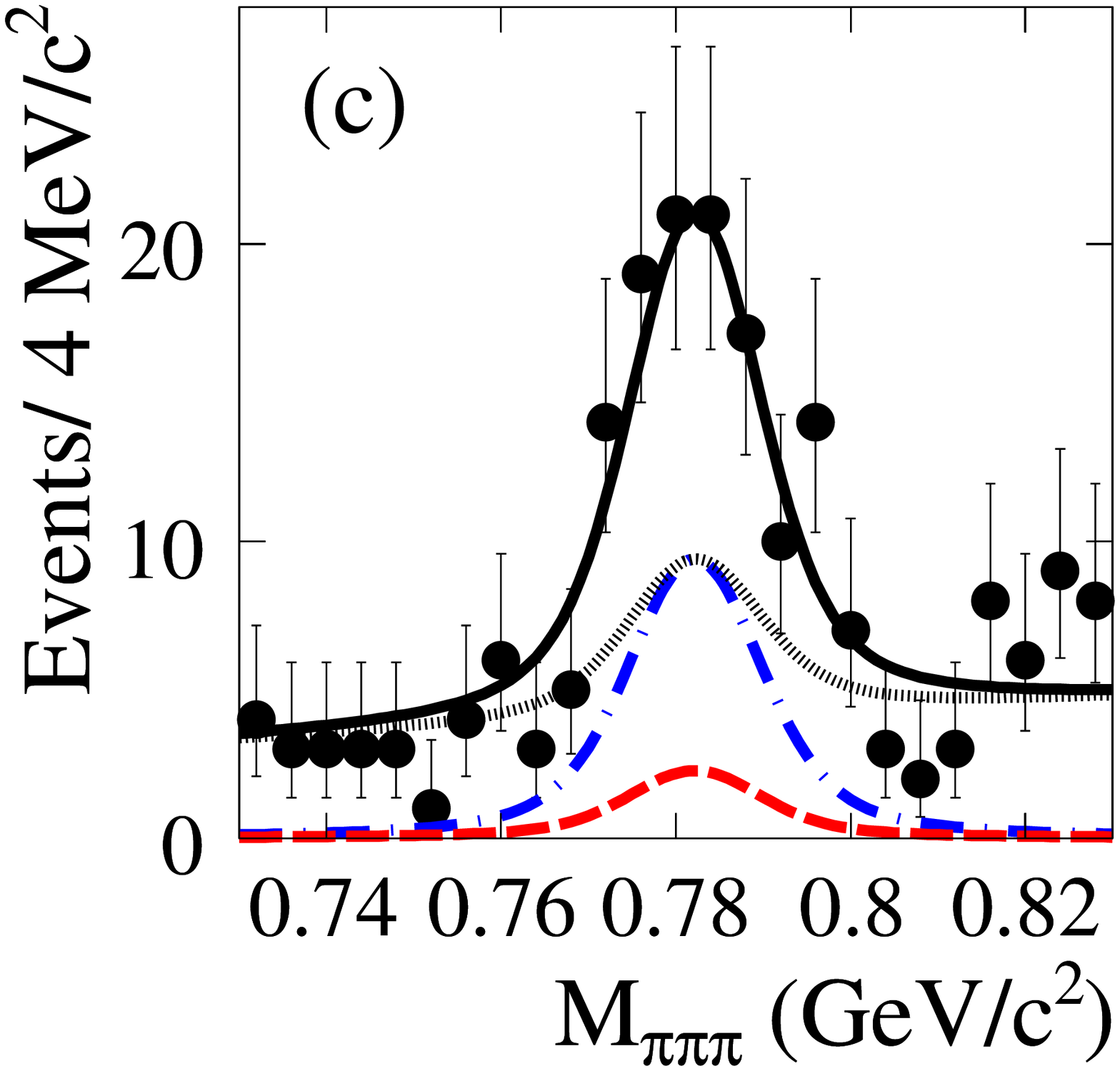}}
\mbox{\epsfxsize=1.68in \epsfbox{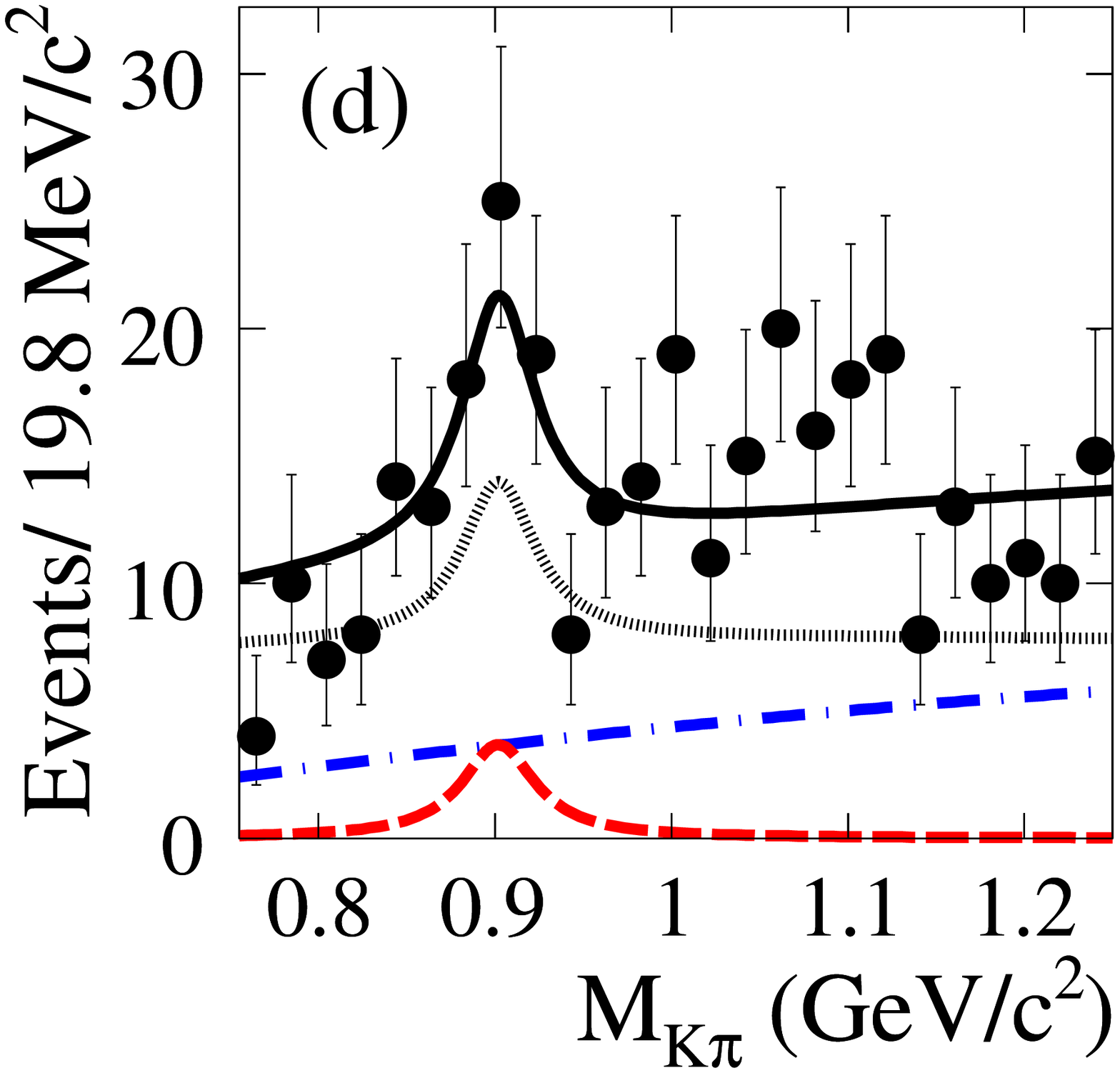}}
\caption{\label{4d fit} Projections of \de (a), \mbc (b), \mOme (c) and \mKstz (d) for events in the signal region of the other three variables. The solid curve is the fit function, the dashed curve is the \BzToOmeKstz component, the dot-dashed curve is the \BzToOmeKpPim component, and the dotted curve is the sum of the \qqbar, \bc and \bsud components. }
\end{figure}

The dominant source of background arises from random combinations of particles in continuum \eeqq events (${q = u,d,s,c}$). To discriminate spherical-like \BBbar events from jet-like \qqbar events, we use event-shape variables, specifically, 16 modified Fox-Wolfram moments combined into a Fisher discriminant, \F~\cite{fisher}. Additional discrimination is provided by \thetab, the polar angle in the CM frame between the \B direction and the negative direction of the positron beam axis. True \B mesons follow a \sigcosthetab distribution, while candidates in the continuum are approximately uniformly distributed in \costhetab. The displacement along the beam axis between the signal \B vertex and that of the other \B, \deltaz, is also used. This variable provides discrimination against continuum events, whose tracks typically have a common vertex. 

Further discrimination against continuum background is achieved through the use of ${b}$-flavor tagging information. The flavor of the \B meson accompanying the signal candidate is identified via its decay products: charged leptons, kaons, and $\Lambda$'s. The Belle tagging algorithm~\cite{tag_algorithm} yields the flavor of the tagged meson, \tagqval, and a flavor-tagging quality factor, \tagr. The latter ranges from zero for no flavor discrimination to one for unambiguous flavor assignment. For signal events, \tagq is usually consistent with the flavor opposite to that of the signal \B, while it is random for continuum events. Thus, the quantity \qrfb is used to separate signal and continuum events, where \fb is the flavor of the signal \B as indicated by the charge of the final state kaon:  \fbCharge for \Bz(\Bzbar).

We use Monte Carlo (MC) simulated signal~\cite{GEANT} and data sideband events (defined as \mbcsbd, \defit) to form \F and obtain the \costhetab, \deltaz and \qrfb distributions. Our signal MC is generated to be 50\% longitudinally polarized (\fL = 0.5). Probability density functions (PDFs) derived from \F, the \costhetab distributions, and the \deltaz distributions are multiplied to form signal (\Lsig) and continuum background (\Lqq) likelihood functions. These are combined to form a likelihood ratio \Lratiodef. We divide the events into six bins of \qrfb and determine the optimum \Lratio selection criteria for each bin by maximizing \fomrhs, where \nsig is the number of signal MC events in the signal region, and \nbkgd is the number of background events estimated to be in the signal region by extrapolating from the data sideband. This optimization preserves 50\% of the signal while rejecting 99\% of the continuum background.

After all selection requirements, 12\% of events in the signal MC have more than one candidate. We choose the best candidate in an event to be the one that minimizes the quantity \bcs. From MC simulation, we find that 8.5\% of signal decays have at least one particle incorrectly identified but pass all selection criteria. We refer to these as ``self-cross-feed" (SCF) events. The remaining signal events, from correctly reconstructed \BzToOmeKstz decays, are referred to as ``true-signal" decays.

We obtain the signal yield using a four-dimensional extended unbinned ML fit to \de, \mbc, \mOme and \mKstz. The likelihood function consists of the following components: true-signal decays, SCF events, non-resonant \Omekpi decays, continuum background (\qqbar), charm \B-decay background (\bc), and charmless \B-decay background (\bsud). For all components, no sizable correlations are found among the fitted quantities. The PDF for event ${i}$ and component ${j}$ is defined as 
\begin{equation} \label{pdf def}
\PDFdefinition.
\end{equation}

For the true-signal, non-resonant, and peaking components of the \qqbar and \bsud backgrounds, the \Kstz and \Ome resonances are modeled with Breit-Wigner functions whose widths are fixed to their PDG~\cite{PDG2004} values. The Breit-Wigner used to describe the \Ome resonance is convolved with a Gaussian of ${\sigma = 5.7~\mathrm{MeV}}$ to account for the detector resolution. This value, along with the means for both resonances and the fraction of peaking \qqbar background events, are obtained from one-dimensional fits to \mKstz and \mOme for events in the data sideband region.

All other parameters are obtained from MC simulation. For the true-signal and non-resonant components, the sum of a Crystal Ball line shape (CBLS)~\cite{cbls} and Gaussian with a common mean is used to describe \de, and the sum of two Gaussians with a common mean is used to describe \mbc. To take into account small differences between the MC and data, the \mbc - \de shapes for the true-signal and non-resonant PDFs are corrected according to calibration factors determined from large \bztodmrhop, \dmtokpp control samples in data and MC. To describe the non-resonant \mKstz component, we use a second-order Chebyshev polynomial. The SCF events are  modeled with non-parametric PDFs using Kernel Estimation~\cite{keys}.

The \mbc PDF for \qqbar background is modeled by a threshold ARGUS~\cite{argus} function. A first-order Chebyshev polynomial is used to model \de and the combinatorial components of \mOme and \mKstz. The PDF for \bc is the product of an ARGUS function for \mbc, and second-order Chebyshev polynomials for \de, \mKstz and \mOme. To model the \bsud background we use a fifth-order Chebyshev polynomial for \de and a Gaussian plus ARGUS function for \mbc. For the non-peaking component of \mOme and \mKstz we use first- and second-order Chebyshev polynomials, respectively.

The following parameters are allowed to vary in our final fit to the data: the true-signal, non-resonant, \bc and \qqbar yields, and the \qqbar PDF parameters describing the \de, \mbc and combinatorial shapes of \mOme and \mKstz. The fraction of SCF events is fixed to be 8.5\% of the signal. The fraction of \bsud events is very small (0.7\%) and thus is also fixed in the fit according to the predictions of MC simulation.

The likelihood function for event ${i}$ is given by
\begin{equation} \label{ll def}
\LFdefinition,
\end{equation}
where ${Y_{j}}$ is the yield of events from component ${j}$ and $N$ is the total number of events in the sample. 

The results of the fit are shown in Fig. \ref{4d fit}. We find strong peaking in \de, \mbc and \mOme, which have shapes consistent with those observed in MC simulations. However, we do not observe a strong \Kstz resonance. Instead, we observe a high density of events in the upper sideband of the \mKstz distribution, which the fit assigns to non-resonant decays. The branching fraction is evaluated using the following quantities: ${\nOmeKstz = 15.1^{+11.1}_{-10.0}}$, the signal yield; \MCeff, the signal efficiency; \PIDeff, an efficiency correction for the charged track selection that takes into account small differences between MC and data; \NBBbarErr, the number of \BBbar pairs produced; and \BRprod, the product of daughter branching fractions.

The sources of systematic error are listed in Table \ref{systematics}. The errors on the PDF shapes are obtained by varying all fixed parameters by ${\pm1\sigma}$. To obtain the error on the SCF and \bsud fractions, we vary the normalizations by $\pm50\%$. The effect of a possible fit bias due to floating the non-resonant \Omekpi normalization is obtained by fitting ensembles of simulated experiments containing varying sets of signal and non-resonant events. We consider the effects of higher \Kstz resonances by calculating the fractional change in our signal yield when we include a PDF for \BzToOmeKstzHigh, whose yield is fixed to 1.0 event, as determined by extrapolating from a higher-mass region. The dependance of the acceptance on \fL is obtained by varying \fL from 0 to 1.0. To obtain the uncertainty on the selection efficiency of the \Lratio requirement, we vary the \Lratio thresholds, and we also calculate the data/MC efficiency ratio for the \bztodmrhop, \dmtokpp control sample.

\begin{table}[t]
\caption{Systematic errors for ${\cal B}$(\BzToOmeKstz).}
\label{systematics}
\vskip0.05in
\begin{tabular}{lrr}
\hline
\textbf {Type} & \multicolumn{2}{c}{ Fractional error  (\%)} \\
            &  $\mathbf{+\sigma}$ & $\mathbf {-\sigma}$  \\
\hline
Track reconstruction efficiency                          &  $4.80$  &  $-4.80$  \\  
\piz reconstruction efficiency                              &         $4.00$         &  $-4.00$  \\  
${K^{\pm} \pi^{\pm}}$ idenfication efficiency   &         $1.33$         &  $-1.33$  \\  
\de PDF shape calibration                                  &         $3.46$         &  $-3.62$  \\  
\mbc PDF shape calibration                               &         $2.12$         &  $-2.19$  \\  
Shape of \Kstz and \Ome PDFs                         &         $3.45$         &  $-3.69$  \\  
Shape of true-signal PDF                                   &         $0.69$         &  $-0.69$  \\  
Shape of \BzToOmeKpPim PDF                        &        $9.02$          & $-5.80$  \\  
Shape of \bc PDF                                                 &         $1.50$         &  $-1.60$  \\  
Shape of \bsud PDF                                            &         $1.02$         &  $-1.09$  \\  
Shape of \qqbar PDF                                           &         $6.65$        &   $-6.16$  \\  
Fraction of \bsud background                            &         $2.59$         &  $-5.04$  \\  
SCF fraction                                                           &         $5.66$        &  $-5.17$  \\  
Possible fitting bias                                              &         $1.50$        & $0.00$  \\  
Effect of higher \Kstz resonances                      &         $1.30$        & $0.00$  \\  
\BzToOmeKstz acceptance                                &          $0.63$       &    $-0.63$  \\  
Longitudinal polarization                                    &          $2.61$       &   $-2.61$  \\  
${\cal R}$ requirement                                         &          $2.80$       &   $-2.80$  \\  
${N_{B\bar B}}$                                                    &           $1.31$       &  $-1.31$  \\  \hline
\textbf{Total}                                                          &   \textbf{16.1}      & \boldmath{$-14.7$}  \\ \hline 
\end{tabular}
\end{table}

Our final result for the branching fraction based on \NBBbar \BBbar pairs is
\begin{equation}
\brfr,
\end{equation}
where the first error quoted is statistical, the second systematic, and the upper limit is taken to be the branching fraction corresponding to 90\% of the total integral of the likelihood function in the positive branching fraction region. The systematic error is included by convolving the likelihood function with a Gaussian having a standard deviation equal to the systematic uncertainty. The statistical significance of the signal, defined as \SqrtNegTwoLnLzLm, where \Lmax (\Lzero) is the value of the likelihood function when \nOmeKstz is allowed to vary (set to 0), is \significanceValue.

To verify the large non-resonant contribution, we fit the background-subtracted \mKstz distribution to extract the signal yield. To obtain this distribution, we bin the data in \mKstz from [0.75, 1.25] \gevcc and, for each bin, perform a two-dimensional extended unbinned ML fit to \de and \mbc. The likelihood function consists of three components: signal + non-resonant, \qqbar + \bc, and \bsud. We use a single PDF to describe the signal + non-resonant component, since their individual shapes are almost identical in \de and \mbc. A single PDF is also used to model \qqbar + \bc, since in several of the bins, the statistics are too low to model them independently. The \de and \mbc PDFs for the signal + non-resonant and \bsud components are identical to those used in the four-dimensional fit, with the exception that here, we do not model the true-signal and SCF events separately. For the \qqbar + \bc PDF, we use a first order Chebyshev polynomial for \de and an ARGUS function for \mbc. We fix the shapes of the signal + non-resonant and \bsud components from MC simulation. In the final fit, we fix the fraction of \bsud, while allowing the other two normalizations, and the \de and \mbc shapes of the \qqbar + \bc PDF, to vary. 

\begin{figure}[]
\mbox{\epsfxsize=3.4in \epsfbox{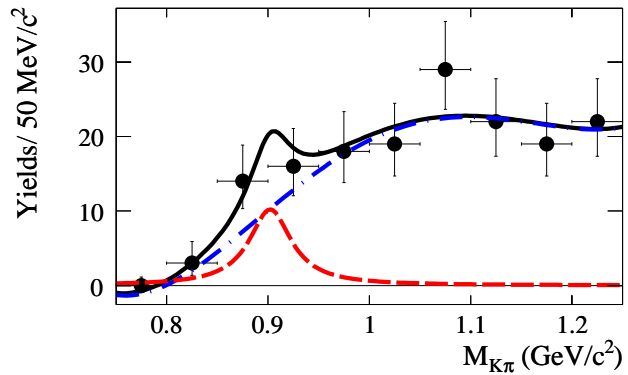}}
\caption{\label{bkgd sub fit} Signal yields obtained from the \de-\mbc distribution in bins of \mKstz for events in the \Ome signal region. The solid curve is the fit function, the dashed curve is the \BzToOmeKstz component, and the dot-dashed curve is the \BzToOmeKpPim component.}
\end{figure}

The results are shown in Fig. \ref{bkgd sub fit}. We perform a ${\chi^{2}}$ fit to this background-subtracted \mKstz distribution. The Breit-Wigner shape is obtained in the same way as for the four-dimensional fit. For the non-resonant component, we use a fourth-order Chebyshev polynomial. In the final fit, we float the non-resonant shape parameters along with the fractional signal yield. We obtain \FracYieldFormulaResultTwoD, which is very similar to the result of the four-dimensional fit: \FracYieldResultFourD.


In summary, we present a measurement of the branching fraction of \BzToOmeKstz decays using \NBBbar \BBbar pairs. The statistical significance of our signal yield is only ${1.6\sigma}$, and thus we set an upper limit of \ul at the 90\% C.L. Our result is in agreement with theoretical estimates~\cite{br_theory}. The limit obtained is below the previous constraint from BaBar~\cite{babar_wKstz_2} by a factor of 1.6. In addition, we observe a large rate for non-resonant \BzToOmeKpPim decays.

We thank the KEKB group for the excellent operation of the
accelerator, the KEK cryogenics group for the efficient
operation of the solenoid, and the KEK computer group and
the National Institute of Informatics for valuable computing
and Super-SINET network support. We acknowledge support from
the Ministry of Education, Culture, Sports, Science, and
Technology of Japan and the Japan Society for the Promotion
of Science; the Australian Research Council and the
Australian Department of Education, Science and Training;
the National Science Foundation of China and the Knowledge
Innovation Program of the Chinese Academy of Sciences under
contract No.~10575109 and IHEP-U-503; the Department of
Science and Technology of India; 
the BK21 program of the Ministry of Education of Korea, 
the CHEP SRC program and Basic Research program 
(grant No.~R01-2005-000-10089-0) of the Korea Science and
Engineering Foundation, and the Pure Basic Research Group 
program of the Korea Research Foundation; 
the Polish State Committee for Scientific Research; 
the Ministry of Education and Science of the Russian
Federation and the Russian Federal Agency for Atomic Energy;
the Slovenian Research Agency;  the Swiss
National Science Foundation; the National Science Council
and the Ministry of Education of Taiwan; and the U.S.\
Department of Energy.


\end{document}

%% file: author-conf2007.tex
\affiliation{Budker Institute of Nuclear Physics, Novosibirsk}
\affiliation{Chiba University, Chiba}
\affiliation{University of Cincinnati, Cincinnati, Ohio 45221}
\affiliation{Department of Physics, Fu Jen Catholic University, Taipei}
\affiliation{Justus-Liebig-Universit\"at Gie\ss{}en, Gie\ss{}en}
\affiliation{The Graduate University for Advanced Studies, Hayama}
\affiliation{Gyeongsang National University, Chinju}
\affiliation{Hanyang University, Seoul}
\affiliation{University of Hawaii, Honolulu, Hawaii 96822}
\affiliation{High Energy Accelerator Research Organization (KEK), Tsukuba}
\affiliation{Hiroshima Institute of Technology, Hiroshima}
\affiliation{University of Illinois at Urbana-Champaign, Urbana, Illinois 61801}
\affiliation{Institute of High Energy Physics, Chinese Academy of Sciences, Beijing}
\affiliation{Institute of High Energy Physics, Vienna}
\affiliation{Institute of High Energy Physics, Protvino}
\affiliation{Institute for Theoretical and Experimental Physics, Moscow}
\affiliation{J. Stefan Institute, Ljubljana}
\affiliation{Kanagawa University, Yokohama}
\affiliation{Korea University, Seoul}
\affiliation{Kyoto University, Kyoto}
\affiliation{Kyungpook National University, Taegu}
\affiliation{Swiss Federal Institute of Technology of Lausanne, EPFL, Lausanne}
\affiliation{University of Ljubljana, Ljubljana}
\affiliation{University of Maribor, Maribor}
\affiliation{University of Melbourne, School of Physics, Victoria 3010}
\affiliation{Nagoya University, Nagoya}
\affiliation{Nara Women's University, Nara}
\affiliation{National Central University, Chung-li}
\affiliation{National United University, Miao Li}
\affiliation{Department of Physics, National Taiwan University, Taipei}
\affiliation{H. Niewodniczanski Institute of Nuclear Physics, Krakow}
\affiliation{Nippon Dental University, Niigata}
\affiliation{Niigata University, Niigata}
\affiliation{University of Nova Gorica, Nova Gorica}
\affiliation{Osaka City University, Osaka}
\affiliation{Osaka University, Osaka}
\affiliation{Panjab University, Chandigarh}
\affiliation{Peking University, Beijing}
\affiliation{University of Pittsburgh, Pittsburgh, Pennsylvania 15260}
\affiliation{Princeton University, Princeton, New Jersey 08544}
\affiliation{RIKEN BNL Research Center, Upton, New York 11973}
\affiliation{Saga University, Saga}
\affiliation{University of Science and Technology of China, Hefei}
\affiliation{Seoul National University, Seoul}
\affiliation{Shinshu University, Nagano}
\affiliation{Sungkyunkwan University, Suwon}
\affiliation{University of Sydney, Sydney, New South Wales}
\affiliation{Tata Institute of Fundamental Research, Mumbai}
\affiliation{Toho University, Funabashi}
\affiliation{Tohoku Gakuin University, Tagajo}
\affiliation{Tohoku University, Sendai}
\affiliation{Department of Physics, University of Tokyo, Tokyo}
\affiliation{Tokyo Institute of Technology, Tokyo}
\affiliation{Tokyo Metropolitan University, Tokyo}
\affiliation{Tokyo University of Agriculture and Technology, Tokyo}
\affiliation{Toyama National College of Maritime Technology, Toyama}
\affiliation{Virginia Polytechnic Institute and State University, Blacksburg, Virginia 24061}
\affiliation{Yonsei University, Seoul}
 \author{K.~Abe}\affiliation{High Energy Accelerator Research Organization (KEK), Tsukuba} 
 \author{I.~Adachi}\affiliation{High Energy Accelerator Research Organization (KEK), Tsukuba} 
 \author{H.~Aihara}\affiliation{Department of Physics, University of Tokyo, Tokyo} 
 \author{K.~Arinstein}\affiliation{Budker Institute of Nuclear Physics, Novosibirsk} 
 \author{T.~Aso}\affiliation{Toyama National College of Maritime Technology, Toyama} 
 \author{V.~Aulchenko}\affiliation{Budker Institute of Nuclear Physics, Novosibirsk} 
 \author{T.~Aushev}\affiliation{Swiss Federal Institute of Technology of Lausanne, EPFL, Lausanne}\affiliation{Institute for Theoretical and Experimental Physics, Moscow} 
 \author{T.~Aziz}\affiliation{Tata Institute of Fundamental Research, Mumbai} 
 \author{S.~Bahinipati}\affiliation{University of Cincinnati, Cincinnati, Ohio 45221} 
\author{A.~M.~Bakich}\affiliation{University of Sydney, Sydney, New South Wales} 
% ...
\collaboration{The Belle Collaboration}

%% file: diagram.tex
\begin{fmffile}{diagram}
\begin{fmfgraph*}(150,70)
     \fmfpen{thick}
    \fmfstraight
    \fmfleft{l2,l1}
    \fmflabel{$\bar{b}$}{l1}
    \fmflabel{$d$}{l2}
    \fmf{phantom,label=$B^0\quad$,l.side=right}{l1,l2}

    \fmfright{r4,r3,r2,r1}
    \fmflabel{$\bar{s}$}{r1}
    \fmflabel{$d$}{r2}
    \fmf{phantom,label=$\quad K^{*0}$,l.side=left}{r1,r2}
    \fmflabel{$\bar{d}$}{r3}
    \fmflabel{$d$}{r4}
    \fmf{phantom,label=$\quad \omega$,l.side=left}{r3,r4}

    \fmf{plain,tension=1.5}{v1,l1}
    \fmf{phantom,tension=1.5}{v1,v3}
    \fmf{plain}{r1,v3}

    \fmf{plain}{l2,v5}
    \fmf{plain}{v5,r4}

    \fmffreeze
    \fmf{boson,label=$W^+$,left=0.8}{v1,v3}
    \fmf{plain,right=0.4,label=$\bar{u},,\bar{c},,\bar{t}$}{v1,v2}
    \fmf{plain,right=0.4}{v1,v2}
    \fmf{plain,right=0.4}{v2,v3}
    \fmf{phantom}{v2,v5}

    \fmffreeze
    \fmf{gluon,label=$g$\quad\quad\quad}{v2,v4}
    \fmf{phantom,tension=0.5}{v5,v4}
    \fmf{plain}{v4,r2}
    \fmf{plain}{r3,v4}
    
    \fmfdotn{v}{4}

\end{fmfgraph*}
\end{fmffile}

%% file: b2kstomega_text.bbl
\begin{thebibliography}{99}
\bibitem{charge_conjugate_text}
Charge-conjugate decays are included unless explicitly stated otherwise.
\end{thebibliography}
